\documentclass[reprint,
superscriptaddress,
 amsmath,amssymb,
prb,
floatfix,
]{revtex4-2}

\usepackage{graphicx}
\usepackage{dcolumn}
\usepackage{mathrsfs}
\usepackage{bm}
\usepackage[T1]{fontenc}
\usepackage{adjustbox}
\usepackage{blindtext}
\usepackage{todonotes}
\usepackage{comment}
\usepackage{xcolor}
\usepackage{soul}

\begin{document}
\setcitestyle{super}
\title{Characterizing the Nonequilibrium Response of FeRh Thin Films using Time-Domain Thermoreflectance (TDTR)}%


\author{Renee M. Harton}
\email[]{hartonr22@gmail.com}
\affiliation{University of Illinois at Urbana-Champaign Department of Materials Science and Engineering,1304 W Green St, Urbana, IL 61801}
\author{Alejandro Ceballos}
\affiliation{Department of Physics, University of California, 366 Physics North
Berkeley, CA, 94720-7300, USA}
\affiliation{Materials Science and Engineering, University of California at Berkeley, Berkeley, CA USA}
\affiliation{Lawrence Berkeley National Lab Materials Science Division, Berkeley, CA USA}
\author{Vivek Unikandanunni}
\affiliation{Department of Physics, Stockholm University, Stockholm, SWE}
\author{Alexander Gray}
\affiliation{Department of Physics, Temple University, Philadelphia, PA USA}
\author{Stefano Bonetti}
\affiliation{Department of Physics, Stockholm University, Stockholm, SWE}\affiliation{Department of Molecular Sciences and Nanosystems, Ca' Foscari University of Venice, Venice ITA}
\author{Peter Kr\"uger}
\affiliation{ Department of Physics, Chiba University, Chiba JPN}
\author{Frances Hellman}
\affiliation{Department of Physics, University of California, 366 Physics North
Berkeley, CA, 94720-7300, USA}
\affiliation{Materials Science and Engineering, University of California at Berkeley, Berkeley, CA USA}
\affiliation{Lawrence Berkeley National Lab Materials Science Division, Berkeley, CA USA}


\date{\today}

\begin{abstract}Time-Domain Thermoreflectance (TDTR) characterization of FeRh throughout its first-order antiferromagnetic (AF) to ferromagnetic (FM) transition shows that the transient reflectance, $\Delta$R(t)/R, strongly depends on the magnetic order of the sample. Using TDTR, which uses optical pulses to induce small temperature excursions, we have found that the $\Delta$R(t)/R of the AF phase exhibits a large negative response, while the response of the FM phase is positive. This magnetic phase sensitivity has allowed us to study the transient response of both the AF and FM phase to the pump pulse excitation and the mixed phase of the material. These results are significant since the ultrafast properties of antiferromagnetic materials and mixed antiferromagnetic and ferromagnetic materials are difficult to detect using other conventional techniques.We have found that the AF phase exhibits a strong subpicosecond signal not observed in the FM phase. The magnetic phase dependence of the sign of $\Delta$R(t)/R is qualitatively explained using the results of ab-initio density functional theory (DFT) calculations. Using the two-temperature model, we found that the change in the thermalization time across the transition is caused by differences in both the electronic heat capacity and the electron-phonon coupling factor of the AF and FM phases. The electron-phonon coupling constant in the AF phase is also determined using the two-temperature model conducted using the NTMpy code package. For the FM phase, we provide boundaries for the magnitude of the electron-phonon coupling factor for the FM phase.These results indicate that TDTR can be used to study the transient properties of magnetic materials that are otherwise challenging to probe.
\end{abstract} 

\keywords{nonequilbrium, FeRh, time-domain, Iron-Rhodium}
\maketitle

\section{\label{sec:level1}Introduction\protect\\}
\begin{figure}[p]
\centering
\includegraphics[width=8.6cm,height=8.6cm,keepaspectratio]{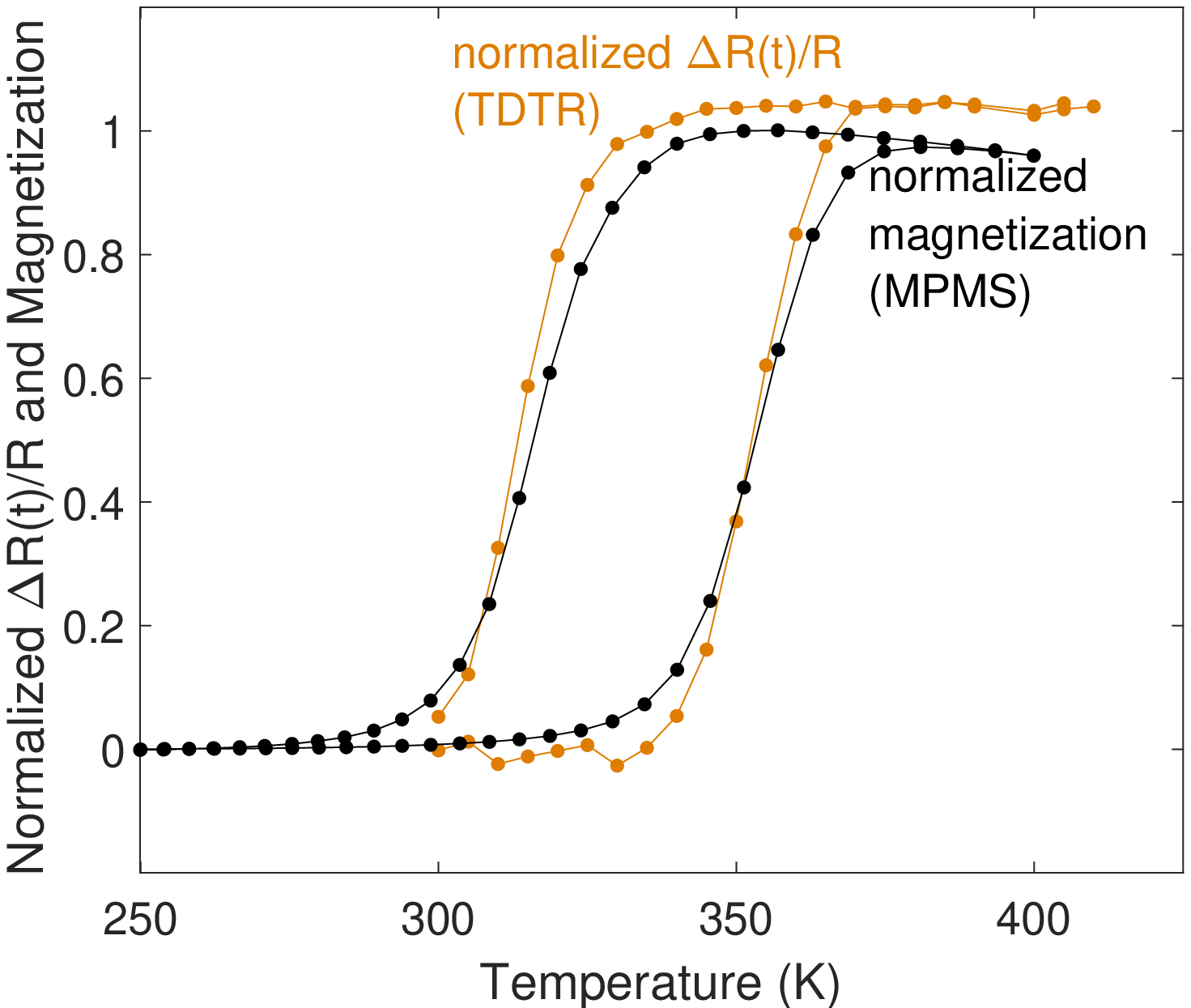}
\caption{\label{fig:MPMSTDTR}\textbf{Temperature dependence of the normalized magnetization of the FeRh film (MPMS signal) and the normalized $\Delta$R(t)/R at a time delay of 1 ps (TDTR signal). Significant hysteresis is observed in both the TDTR and MPMS signals. Both the TDTR and MPMS signals are normalized for relative comparison. Since the MPMS detects the magnetization in the absence of excitation by an optical pump pulse,} the overlap of the TDTR and MPMS signals suggest that for a given base temperature, the FM phase fraction at a time delay of one picosecond is effectively equivalent to the equilibrium magnetization at the base temperature.}
\end{figure}

 Iron-rhodium alloys, Fe$_{x}$Rh$_{1-x}$, with Fe compositions, x, between 0.48$<$x$<$0.52, exhibit a first-order magnetic phase transition near room temperature.\cite{lewis_coupled_2016} Although this phase transition was first observed by Fallot et al. in 1938,\cite{fallot_les_1938} it has long been investigated by researchers due to its advantageous properties such as large magnetoresistance\cite{algarabel_giant_1995} and magnetostriction.\cite{ibarra_giant_1994} In addition, its proposed technological applications, for example, heat-assisted magnetic recording are of interest.\cite{thiele_ferhfept_2003,huang_approaching_2014} This attention has resulted in a rich body of research focused on the structural and magnetic properties of FeRh alloys.\newline\indent The magnetic phase transition occurs across a range of temperatures and exhibits hysteresis which is dependent on both composition and microsctructure.\cite{cher_compositional_2011,ceballos_effect_2017} For temperatures below the transition, the magnetic order of the material is antiferromagnetic (AF); above the transition region, the material is ferromagnetic (FM). For temperatures within the AF/FM transition region, the sample exhibits a mixed phase of AF and FM domains.\cite{baldasseroni_temperature-driven_2015,keavney_phase_2018} Near and throughout the transition region, FeRh crystallizes in CsCl structure; as the sample undergoes the AF/FM transition with increasing temperature, the lattice constant increases by $\approx$\ 0.1$\%$.\cite{swartzendruber_ferh_1984} \newline\indent The intriguing nature of the FeRh AF/FM transition has motivated researchers to investigate its origin and to examine the different properties of the AF and FM phases. These studies have led to a thorough exploration of the different properties of the AF and FM phase of FeRh. First principle calculations using density functional theory show that the density of states (DOS) of the AF phase exhibits a pseudogap of approximately 0.5 eV near the Fermi energy.\cite{bennett_spectral_2019} For the FM domains, this pseudogap in the electronic DOS is not observed. These results are supported by Hall effect measurements which determine that the Hall coefficient decreases by almost an order of magnitude as the number and size of the FM domains grows with an increase in temperature in the transition region.\cite{de_vries_hall-effect_2013} The source of this difference in Hall coefficient is an increase in the carrier density concomitant with an increase in the density of states near the Fermi energy in the FM phase relative to the AF phase. Another experimental result supporting the pseudogap of the AF phase was found by Pressacco et al.\cite{pressacco_laser_2018} Using X-Ray Photoemission Spectroscopy, Pressacco et al. demonstrated an increase in the electron spectral weight with the formation of FM domains. These results show that as the FM phase grows in size with an increase in the base temperature of the sample, the density of states near the Fermi energy also increases. One might wonder how the presence of the pseudogap in the AF phase and the absence of this feature in the FM phase affects the response of FeRh to an optical pump pulse. \newline\indent The response of FeRh to optical pump pulse excitation has intrigued researchers for decades. Pump-probe techniques, such as Time-Resolved Magneto-Optic Kerr Effect(TR-MOKE) magnetometry, Time-Resolved x-ray diffraction, and Time-Resolved photoelectron spectroscopy have been used to study the transient response of FeRh. From these studies, the onset of FM order has been observed on both the subpicosecond\cite{ju_ultrafast_2004,thiele_spin_2004,pressacco_subpicosecond_2021} and picosecond timescales.\cite{mariager_structural_2012,pressacco_laser_2018,li_ultrafast_2022} As a result, the limiting timescale of the AF/FM transition remains a source of debate.\newline\indent Although the timescale of the FeRh AF/FM transition has been reported using several detection methods, these measurements used large pump fluences. Moreover for the referenced studies, the pump fluence was systematically increased to induce the transition.\cite{ju_ultrafast_2004,bergman_identifying_2006} These measurements identified a threshold pump fluence required to drive the AF/FM transition. Since the nonequilibrium response of metals depends on pump fluence,\cite{mueller_relaxation_2013} it is challenging to decouple the effects caused by changes in phase fraction from those caused by altering the fluence of the pump. In the reported studies, we used time-domain thermoreflectance (TDTR) to determine the response of FeRh thin films to pump excitation. For these measurements, the temperature change per pulse of the pump beam was 2K.  The width of the AF/FM transition for comparable FeRh films is approximately 70 K.\cite{yu_thickness-dependent_2019} Since the temperature change induced by a single pulse used for this measurement is less than 3$\%$ of the width of the AF/FM transition, we were able to study dynamics of FeRh for temperatures within the AF and FM regions as well as throughout the AF/FM transition region.\newline\indent Although TR-MOKE is able to detect the FM phase, the dynamics of the AF phase have been difficult to measure using these methods. Recently, quadratic MOKE has been used to study the magnetization dynamics of the metallic antiferromagnet, Fe$_{2}$As.\cite{yang_magneto-optic_2019} The analysis of this method requires that the components of the dielectric function tensor that are linear and quadratic in magnetization are zero. FeRh exhibits a remnant magnetization in the AF phase due to a residual FM component and a mixed phase throughout the AF/FM transition. Thus, the nonequilibrium response of the AF phase of FeRh as well as the response of the mixed phase is difficult to isolate using quadratic MOKE. Our results show that the transient reflectance can be used to detect the subpicosecond response of AF domains in the transition region when FM domains are present, which to the best of our knowledge, cannot be detected using other conventional methods. \newline\indent The transient reflectance, $\Delta$R(t)/R, is commonly used to study the electronic and structural properties of materials on the subpicosecond timescale. Thus, the transient reflectivity of FeRh across the AF/FM transition has been measured, previously.\cite{ju_ultrafast_2004,thiele_spin_2004,bergman_identifying_2006} The  $\Delta$R(t)/R was used to deduce the response of the lattice to pump pulse excitation. For these measurements, large pump fluences are used to drive the system through the AF/FM transition.\cite{thiele_spin_2004,ju_ultrafast_2004} Since the transient reflectance depends on the electronic bandstructure of a material, one might consider if the transient reflectance of FeRh is sensitive to the magnetic phase of the material.\cite{diroll_broadband_2020}This question is the focus of this paper.
\newline\indent We have studied the temperature dependence of the transient reflectance of FeRh using time-domain thermoreflectance. The transient reflectance exhibits a significant dependence on temperature. To associate these changes in the temperature dependence of the transient reflectance with the FeRh AF/FM phase, we have incorporated SQUID magnetometry results into our study. Comparing the magnetic hysteresis shortly following pump pulse excitation to that observed in the absence of pump excitation, we have displayed the temperature dependence of the magnetization in the absence of pump excitation (SQUID magnetometry) with the temperature dependence of the transient reflectance at a time delay of one picosecond(TDTR) in Figure \ref{fig:MPMSTDTR}. The overlap in the temperature dependence of these two signals supports the conclusion that the AF/FM phase fraction is not significantly changed by pump-pulse excitation. Thus, the measured transient reflectance at a given base temperature is a property of the AF/FM phase fraction measured in the absence of pump excitation. Using the AF/FM phase-sensitivity of this technique, we have studied the dependence of the system response of FeRh on the relative AF/FM phase fraction for time delays shorter than the thermalization time of the electrons of the system. \newline\indent It is well known that the reflectance of FeRh is dependent on the magnetic order of the material. The relative magnitude of the reflectance of the AF phase and FM phase depends on the wavelength of the probe beam.\cite{saidl_investigation_2016,bennett_spectral_2019} However, the results of this study are significant because we are able to detect the AF phase on the subpicosecond timescale and thus observe how it responds to pump pulse excitation for base temperatures below and throughout the AF/FM transition. Using the same technique, we can then compare these results to the response of the FM phase. To analyze these results, we have used density functional theory and the two-temperature model to model the response of the system to pump excitation.
\section{\label{sec:level2}Experimental Methods\protect\\} Using TDTR, we measured the transient reflectance of Pt(3 nm)/Fe$_{49}$Rh$_{51}$(19 nm)/MgO(100) sample at different base temperatures of the sample. The epitaxial FeRh thin film was grown using dc magnetron sputtering deposition with a single equiatomic FeRh target. The chamber base pressure was 8$\times$10$^{-8}$ torr, with a growth pressure of 2 mtorr of Ar. During the FeRh film deposition, the substrate temperature was maintained at 873 K. The thickness of the deposited FeRh film was 19 nm. Following the film deposition, a Pt capping layer was deposited at room temperature to prevent oxidation of the FeRh film. Further details on the sample growth and characterization can be found in the cited literature.\cite{ceballos_effect_2017,baldasseroni_effect_2014}The sample compositions were determined using Rutherford backscattering spectrometry. The temperature dependence of the AF/FM transitions of the sample was characterized using a Magnetic Property Measurement System (MPMS).\newline\indent For the MPMS measurements, the sample was configured such that the applied magnetic field was parallel to the sample plane. To align remnant FM domains in the sample, we applied a magnetic field during the temperature-dependent measurements. We determined the required magnitude of the applied field by conducting magnetic field-dependent measurements at room temperature. The minimum field needed to overcome the domain pinning energy was 900 Oe. This magnitude was small enough that both the diamagnetic response and the shift in the temperature of the AF/FM transition were negligible.\cite{maat_temperature_2005}
\newline\indent For the TDTR measurements, a mode-locked Titanium:Sapphire ultrafast laser with a pulse repetition rate of 80 MHz was used. The central wavelength of the generated pulses was set to 783 nm. The pump and probe beams were separated using a polarizing beam splitter(PBS) and thus were cross-polarized. The beams were spectrally separated using 785 nm RazorEdge ultrasteep short-pass edge filter with a cutoff frequency of 785 nm for the pump beam and a 785 nm RazorEdge ultrasteep long-pass edge filter for the probe beam.\cite{kang_two-tint_2008}\newline\indent For the described measurements, the powers of the pump and probe beams were 7 mW and 3 mW, respectively. These beams were focused onto the sample using a 5x microscope objective lens with a 10.6 $\mu$m spot size. The temperature excursion per pulse of the pump and probe beams were approximately 2 K and 1 K, respectively. The temperature excursion per pulse is approximated as (P(1-R))/(f$_{\text{rep}}\pi \text{w}_{\text{o}}^2 \text{hC}$), where f$_{\text{rep}}$,$\text{w}_{\text{o}}$, h and C are the pulse repetition rate, beam spot size, optical absorption depth, and the heat capacity of the sample respectively. The transmitted power is determined by measuring the power of the reflected beam from the sample. The steady-state heating of the pump and the probe beams was approximately 1 K and 0.8 K, respectively.\cite{braun_steady-state_2018}\newline\indent The pump-induced effects on the sample were measured using lock-in detection. The pump beam was modulated at 11 MHz using an electro-optic modulator. Double-modulation was used to reduce the radio-frequency coherent pickup.\cite{jiang_tutorial_2018} Double-modulation was achieved using a computer-based lock-in audio-frequency detector while the probe beam was modulated at a rate of 200 Hz using a mechanical chopper. The time delay between the pump and probe pulses was achieved by altering the relative path length of the pump and probe beam path using a retroreflector along the pump beam path. The pump and probe beams were both normally incident on the sample.\newline\indent To ensure uniform heating the FeRh sample was mounted onto a SiO$_{2}$(300 nm)/Si substrate using Ag paste. Measurements were conducted using an Instec heating and cooling stage to control the base temperature of the sample. We determined the temperature range for the TDTR measurements using the results of the magnetic hysteresis characterization of the sample. The range of base temperatures used for this study was 300 K-410 K. The pump-probe correlation function was determined by a cross-correlation measurement using the pump and the probe beams. For this measurement, we used a Thorlabs DET25K GaP detector. \newline\indent The ab-initio density functional theory calculations were conducted using the Vienna Ab intio Simulation package (VASP). For these calculations, the FeRh alloy is assumed to have perfect structural and chemical ordering. A more detailed description of the methods used in these calculations can be found in the work conducted by Gray et al.\cite{gray_electronic_2012} The two-temperature model calculations were conducted using the open-source NTMpy code package.\cite{alber_ntmpy_2021} In these simulations, the sample was modeled as an FeRh layer on an MgO substrate. The resolution of the grid was increased at the FeRh/MgO interface until the solutions converged.
\section{\label{sec:level3}Experimental Results\protect\\}
\begin{figure}[p]
\centering
\includegraphics[width=8.6cm,height=8.6cm,keepaspectratio]{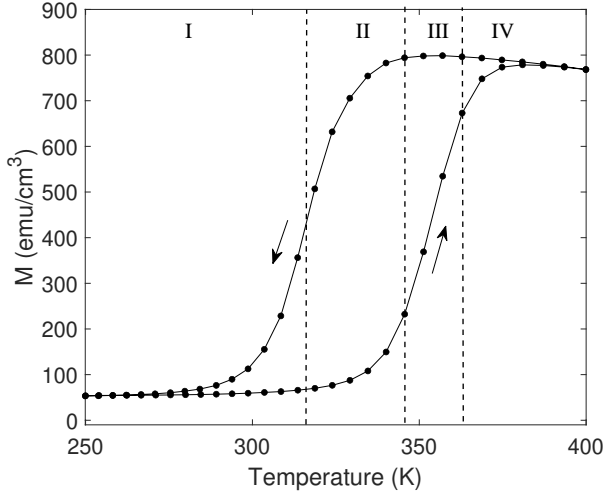}
\caption{\label{fig:mag19nm}\textbf{Temperature-dependent magnetization measurements of the Pt(3 nm)/FeRh(19nm)/MgO sample with an applied magnetic field of 900 Oe}. The temperature range is divided into four regions, which correspond to the phase of the FeRh sample.} In temperature region I, the majority of the sample consists of antiferromagnetic (AF) domains. In addition, there is a remnant ferromagnetic phase with a magnetization of approximately 59 emu/cm$^{3}$. The sample undergoes the AF/FM transition in regions II and III. During this temperature range, the sample exhibits a mixed phase. In temperature region IV, the majority of the sample is FM. As a result, the magnetization decreases as the sample approaches the Curie temperature of FeRh at 660 K.\cite{Zakharov2003MagneticAM}
\end{figure}

\begin{figure}[p]
\centering
\includegraphics[width=17.2cm,height=8.6cm,keepaspectratio]{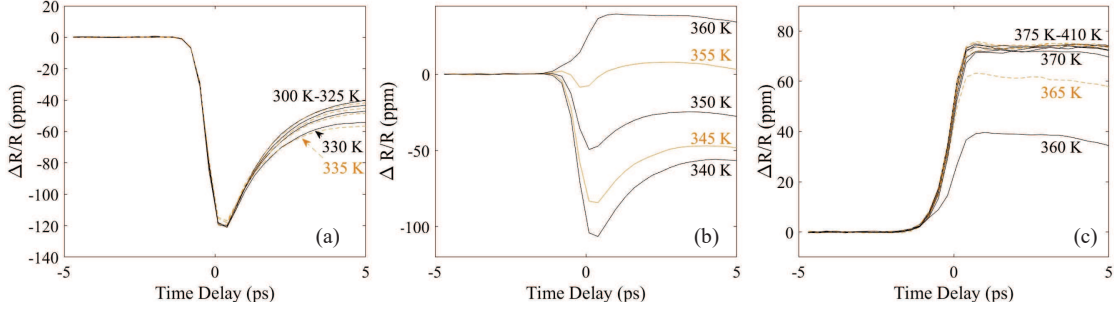}\caption{\label{fig:TDTR}\textbf{Time dependence of the transient reflectance ($\Delta$R(t)/R) of the FeRh sample at different sample base temperatures. The $\Delta \text{R(t)/R}$ is displayed in the majority AF phase (a), AF/FM transition region (b), and the majority FM phase (c).} In both the majority AF and FM phases, the time dependence of $\Delta$R(t)/R does not exhibit a significant dependence on the sample base temperature. However, the transient response of the AF phase (a) shortly following pump excitation is not detected in the FM phase (c). These results highlight differences in the nonequilibrium responses of FeRh in each phase. In the AF/FM transition region (b) when FeRh exhibits a mixed phase, the transient response of $\Delta$R(t)/R characteristic of the AF phase is identified. As expected, this feature of the transient signal decreases in magnitude as the FeRh enters the majority FM region with an increase in the base temperature of the sample.}
\end{figure}

The magnetic hysteresis of the FeRh sample is displayed in Figure \ref{fig:mag19nm}. The temperature dependence of the magnetization of FeRh can be divided into four regions; each region includes increasing and decreasing temperatures. These regions were determined using the results of Baldasseroni et al. where the slope of the magnetization with respect to temperature was associated with different mechanisms for the AF/FM transition.\cite{baldasseroni_temperature-driven_2012}In temperature Region I, the sample is majority AF, with a remnant FM phase. In temperature regions II and III, the sample undergoes the AF/FM transition. For temperatures within these regions, the sample employs two mechanisms to undergo the AF/FM transition: nucleation and domain growth.  In Region II there is an increase in the number of FM domains by nucleation at sites within the AF background. In Region III, the number of FM domains remains roughly the same while the nucleated FM domains increase in size. The growth of the FM domains is reflected in Figure \ref{fig:mag19nm} by the change in the slope of the magnetization with respect to temperature.  In Region IV, the sample is majority FM. As the temperature is further increased, the magnetization decreases as the sample approaches the Curie temperature of FeRh (T$_{\text{C}}$=660 K).\cite{Zakharov2003MagneticAM} In temperature Region I, when the sample consists of mostly AF domains, a remnant magnetization of 59 emu/cm$^3$ is observed. This nonzero magnetization below the onset of the transition can be explained by residual FM domains, which have also been observed in similar FeRh samples.\cite{drozdz_perpendicular_2020,lu_effect_2013,fan_ferromagnetism_2010} The measured remnant magnetization accounts for approximately 7$\%$ of the total magnetization in the FM phase. Therefore, the majority of the sample exhibits AF order below the AF/FM transition region and undergoes the AF/FM transition.
 \newline\indent The time dependence of the transient reflectance, $\Delta$R(t)/R, for temperatures within Region I are displayed in Figure \ref{fig:TDTR}a. In this region, when the sample consists of a majority of AF domains, a negative peak in the transient reflectance is detected, reaching a minimum value at a time delay of approximately 0.7 ps. Since this time delay is equivalent to the full-width half maximum (FWHM) of the temporal correlation of our setup, we conclude that the rise time of this peak is on the subpicosecond timescale. Additionally, the fall time of this peak does not exhibit a noticeable dependence on temperature. \newline\indent At the onset of the AF/FM transition region (Region II and III) when the sample exhibits a mixed phase, the transient signal characteristic of the AF phase is observed. The $\Delta$R(t)/R for temperatures within these regions is displayed in Figure \ref{fig:TDTR}b. The peak magnitude decreases as the base temperature of the sample increases. At the center of the transition region, the peak magnitude of $\Delta$R(t)/R becomes positive and continues to increase as the base temperature is further increased.
\newline\indent In addition to the peak magnitude, the fall time of the transient signal also depends on temperature. At the onset of the AF/FM transition (Region II), the fall time of the peak is equivalent to that of the AF phase. However at higher base temperatures (Region III), the fall time decreases with increasing sample base temperature. Moreover, as the base temperature of the sample reaches Region IV, the fall time of the transient peak is no longer detected. Since the sample consists of a majority of FM domains in this region, we conclude that the absence of the transient signal is a feature of $\Delta$R(t)/R of the FM domains of the sample.  Similar behavior is observed for the $\Delta$R(t)/R for decreasing temperatures. The dependence of the sign of $\Delta$R(t)/R on the AF and FM phase, Region I and Region IV, suggests that $\Delta$R(t)/R depends on AF/FM phase fraction.\newline\indent In addition to the dependence of the magnitude of $\Delta$R(t)/R on the phase of the material, the results of our measurement exhibit a dependence of the fall time of $\Delta$R(t)/R on the relative number of AF and FM domains in the sample. \section{\label{sec:level4}Phenomenological Model\protect\\}
In metals, it has been demonstrated that hot electrons are excited above the Fermi energy upon excitation by an optical pump pulse.\cite{stohr_magnetism_2006} As a result, the electron distribution can no longer be described using Fermi-Dirac statistics. As time progresses, the hot electrons lose energy through electron-electron scattering until thermal equilibrium is reached with the electrons near the Fermi surface. \newline\indent Shortly following the thermalization time of the electron distribution, the hot electrons diffuse through the sample where they lose energy by scattering with phonons in the lattice. These electron-phonon scattering events raise the temperature of the lattice until the electrons and phonons reach thermal equilibrium. After thermalization is achieved, electrons diffuse from the surface toward the substrate.\cite{hohlfeld_electron_2000}  Using the temperature dependence of $\Delta$R(t)/R, we modeled the response of the system to pump pulse excitation before the electron and lattice temperatures thermalize. We used the two-temperature model and linear systems techniques.\subsection{\label{sec:level4TTM}Two-Temperature Model\protect\\} The two-temperature model utilizes two thermally coupled reservoirs, one for the electrons and another for the lattice, each with its own temperature.\cite{bennemann_nonlinear_1998} Although FeRh exhibits magnetic ordering, the intrinsic Gilbert damping parameter of both the AF and FM phase is low, 0.0024 and 0.0031, respectively.\cite{wang_spin_2020} As a result, the electron and spin reservoirs are weakly coupled. Consequently, the response of the FeRh system can be approximated using two temperatures, one for the electron reservoir and another for the lattice reservoir (two-temperature model).\cite{barker_higher-order_2015}\newline\indent Using the two-temperature model, the time evolution of the electron and lattice temperatures (T$_{E}$(t) and T$_{L}$(t)) are described by the following coupled equations: \begin{equation}\label{eq:TTM1}\begin{split} \text{C}_\text{E}\text{(T)}\frac{\text{dT}_{\text{E}}\text{(t)}}{\text{dt}}&=-\text{G}_{\text{EP}}(\text{T})[\text{T}_{\text{E}}\text{(t)}-\text{T}_{\text{L}}\text{(t)}]+\text{S(t)} \\ \text{C}_\text{L}\text{(T)}\frac{\text{dT}_\text{L}\text{(t)}}{\text{dt}}&=\text{G}_{\text{EP}}\text{(T)}[\text{T}_{\text{E}}\text{(t)}-T_{\text{L}}\text{(t)}]
 \end{split}\end{equation} where S(t) and G$_{\text{EP}}$(T) are the absorbed energy from the optical pulse and the electron-phonon coupling factor, respectively. C$_{\text{E}}$(T) and C$_{\text{L}}$(T) are the electronic and lattice heat capacities, respectively. For our measurements, $\Delta$T$_{\text{E}}<< $T$_{0}$, where T$_{0}$ is the base temperature of the sample. The small magnitude of the temperature excursion of the electron temperature in relation to the base temperature is confirmed by the similar temperature dependence of $\Delta$R/R(t$_{\text{d}}$=1 ps) and the MPMS data (Figure \ref{fig:MPMSTDTR}). These results can be explained by the small pump fluence used during these measurements ($\approx$0.1 mJ/cm$^2$). The small temperature excursion in response to the optical pump pulse results in a small change in the electron and lattice temperatures following excitation by the pump pulse. Thus, the weak perturbation approximation is appropriate to solve the coupled differential equations displayed in Equation \ref{eq:TTM1}.\cite{prasankumar_optical_2011} This approximation assumes that C$_{\text{E}}$(T), C$_{\text{L}}$(T) and G$_{\text{EP}}$(T) are effectively constant throughout the calculation of the electron and lattice temperatures. Within the weak perturbation regime, T$_{\text{E}}$(t) is exponential with the following form, T$_{E}$(t)=Ae$^{-bt}$, where \begin{equation}\label{eq:b}b=\frac{\text{G}_{\text{EP}}(C_{E}+C_{L})}{C_{E}C_{L}}.\end{equation} \newline\indent Although the fast exponential decay is detected in the AF phase,  it is not resolved in the FM phase. These results highlight the phase dependence of the exponential decay rate and thus the thermalization time of FeRh. The phase dependence of the thermalization time in FeRh is supported by earlier studies reporting the phase dependence of the electronic and lattice heat capacities of FeRh.\cite{cooke_thermodynamic_2012}The values of the electronic and lattice heat capacities for the AF and FM phases are displayed in Table \ref{tab:partable}.
 \begin{table}
\begin{tabular}{||c| c |c| c| c||} 
 \hline
Phase&$\Theta_{\text{D}}(K)$ &$ \gamma$ & C$_{L}$ (300 K) &C$_{L}$(390 K) \\ [0.5ex] 
 \hline\hline
 AF&393 & 216.9 & 2.78$\times10^{6}$ &2.89$\times10^{6}$ \\ 
 \hline
 FM& 340 & 514.3 & 2.69$\times$ 10$^{6}$ & 2.74$\times10^{6}$ \\ \hline  
\end{tabular}\caption{\label{tab:partable} Electronic heat capacity [$\gamma$T] and Lattice Heat Capacity [C$_{L}$(T)] in majority AF phase (300K) and majority FM phase (390 K). The units of the electronic heat capacities and the lattice heat capacities are  $\frac{\text{J}}{\text{m}^{3}\text{K}^{2}}$ and $\frac{\text{J}}{\text{m}^{3}\text{K}}$, respectively. The Debye temperatures were reported by Cooke et al.\cite{cooke_thermodynamic_2012} To calculate the lattice heat capacities at both 300 K and 390 K, we determined C$_{\text{v}}$(T) using the Debye model. The lattice heat capacity [C$_{\text{L}}$(T)] was determined using C$_{\text{v}}$(T)-$\gamma$T, where C$_{V}$(T) is the heat capacity a constant volume.}
\end{table}
These values show that the reduced heat capacities of the AF and FM phases have the following relationship:
\begin{equation}\label{eq:TTM} \left[\frac{\text{C}_{\text{E}}+\text{C}_{\text{L}}}{\text{C}_{\text{E}}\text{C}_{\text{L}}}\right]_{\text{FM}}<\left[\frac{\text{C}_{\text{E}}+\text{C}_{\text{L}}}{\text{C}_{\text{E}}\text{C}_{\text{L}}}\right]_{\text{AF}}.
\end{equation} As b$_{FM}$ is larger than b$_{AF}$, Equation \ref{eq:b} yields G$_{\text{EP}}\text{(FM)}>$G$_{\text{EP}}\text{(AF)}$. \newline\indent Since the fast exponential decay of the transient reflectance of the FM phase could not be resolved by our setup, we used the system resolution limit to determine the maximum decay rate that our system could detect. This value serves as a maximum decay rate for the FM system.  Since the resolution of the TDTR setup is 0.77 ps, the minimum decay rate is 1.3$\times$10$^{12}$s$^{-1}$. Thus, using Equation \ref{eq:b}, the maximum G$_{\text{EP}}$ at T= 390 K is 2.43$\times10^{17}$ W/m$^{3}$K.\newline\indent In the AF phase, the fast decay rate was detected by our system. As a result, we were able to determine the G$_{\text{EP}}$ of the AF phase using the two-temperature model. This technique required that we fit the $\Delta$R(t)/R using a weighted average of the electron and lattice temperatures calculated by the NTMpy code for a given $G_{\text{EP}}$. To fit the transient reflectance using the weighted average of the electron and lattice temperatures, we first fit the electron and lattice temperatures generated by the NTMpy code using linear systems techniques.\subsection{\label{sec:level4LS}Linear Systems Model\protect\\} 
\begin{figure}[p]
\centering
\includegraphics[width=17.2cm,height=8.6cm,keepaspectratio]{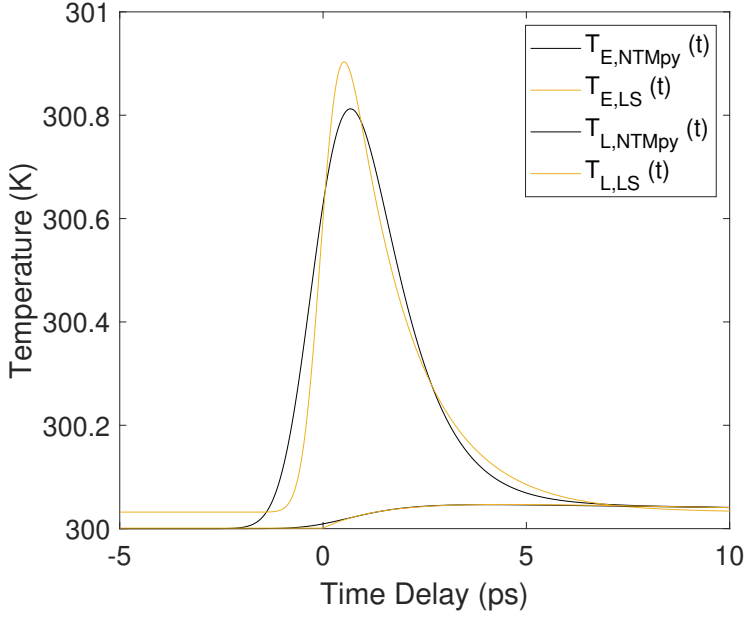}
\caption{\label{fig:fits}\textbf{Fitted time-dependence of electron temperature [T$_{\text{E,NTMpy}}$(t)] and lattice temperature [T$_{\text{L,NTMpy}}$(t)] calculated using two-temperature model.} T$_{\text{E,NTMpy}}$(t) and T$_{\text{L,NTMpy}}$(t) were calculated using the NTMpy python package with an electron-phonon coupling factor of 7.11$\times$10$^{16}$ W/m$^{3}$K at 300 K. T$_{\text{E,LS}}$(t) and T$_{\text{L,LS}}$(t) were determined using linear systems techniques. A weighted sum of T$_{E,LS}$(t) and T$_{L,LS}$(t) is used to fit the transient reflectance ($\Delta$R(t)/R) at 300K in Figure \ref{fig:gepfit}.}
\end{figure}

\begin{figure}[p]
\centering
\includegraphics[width=8.6cm,height=8.6cm,keepaspectratio]{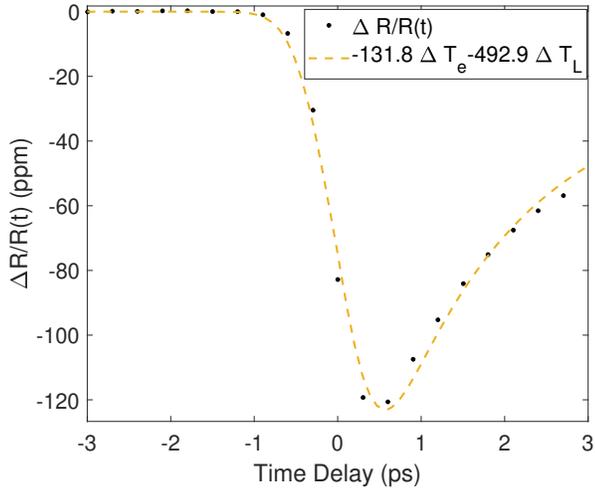}
\caption{\label{fig:gepfit} Fit of the transient reflectance ($\Delta$R(t)/R) of FeRh at room temperature using a weighted sum of the electron temperature (T$_{E}$(t) and T$_{L}$(t) calculated using the NTMpy code. An electron-phonon coupling factor G$_{EP}$(300 K) of 7.11$\times10^{16}$ W/m$^{3}$K.} 
\end{figure}

Since the magnitude of the pump-induced temperature excursions was only 1$\%$ of the base temperature of the sample, linear systems techniques could be used to fit the calculated electron and lattice temperatures. For time delays shortly following pump excitation (t$_{\text{delay}}\approx$ 1-2 ps),  the system response, g(t), was determined by convolving the pump-probe correlation function, C$_{\text{pp}}$(t), with the impulse response of the system, h$_{\text{E}}$(t). The system response can be represented using an expression of the following form, \begin{equation}\label{eq:2}g(t)=Q\times \text{C}_{\text{pp}}(t)\ast \text{h}_{\text{E}}(t),\end{equation} where $Q=\sqrt{\text{S}_{\text{2p}}\times\text{G}}$. S$_{\text{2p}}$ is the two-photon quantum efficiency and G is the impedance gain of the detector.\cite{shin_simple_2016} The time dependence of the system response was modeled using a decaying exponential,\begin{equation}\label{eq:3}\text{h}_{\text{E}}\text{(t)}=\text{a}_{e}e^{-\text{b}_{\text{E}}t},\end{equation} where a$_{\text{E}}$ and b$_{\text{E}}$ are fitting parameters. Equation \ref{eq:3} was used to fit the electron temperature calculated using the two-temperature model. The calculated lattice temperature was fit using the following functional form:\begin{equation} \text{g}_{\text{L}}\text{(t)}=(1-\text{e}^{-\text{b$_{\text{L}}$t}})\left(\frac{\text{a}}{(1+\text{t}/\text{t}_{\text{diff}})^{1/2}}\right)\text{H}\text{(t)}\end{equation} where a,b, and t$_{\text{diff}}$ are fitting parameters and H(t) is the heaviside step function. The fitted time-dependence of the electron and lattice temperatures are displayed in Figure \ref{fig:fits}.\newline\indent Using the NTMpy code to determine the electron and lattice temperatures for a given electron-phonon coupling factor along with the linear systems model to fit the weighted sum of the electron and lattice temperatures to the measured $\Delta$R(t)/R, we have determined G$_{\text{EP}}$(300 K)=7.11$\times$10$^{16}$W/m$^{3}$K. These fit results and the associated T$_{E}$(t) and T$_{L}$(t) in the inset are displayed in Figure \ref{fig:gepfit}.  \section{\label{sec:level5}Discussion\protect\\}
\begin{figure}[p]
\centering
\includegraphics[width=17.2cm,height=8.6cm,keepaspectratio]{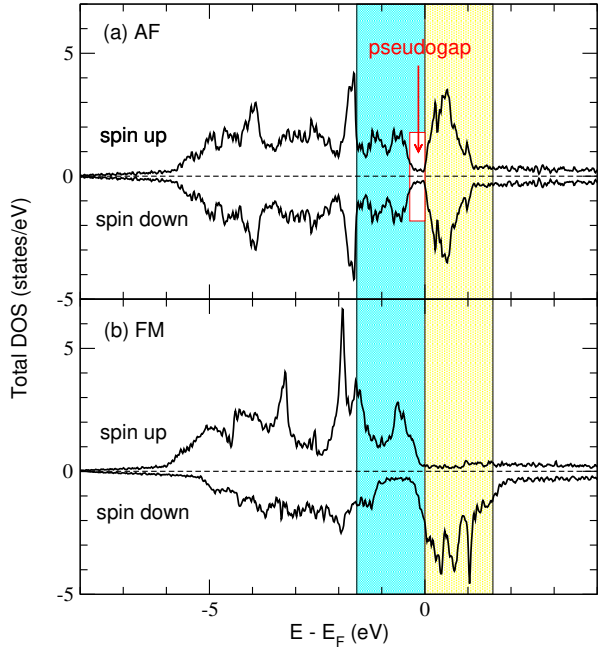}
\caption{\label{fig:AFDOS} Density of states (DOS) of FeRh in (a) AF phase and (b) FM phase
calculated in DFT.
The Fermi level (E$_F$) is put to zero. Spin-up DOS is plotted in blue and
spin down DOS in red.
The area shaded in yellow shows the states that may be excited with the
1.58 eV photons used in the experiment.
The area shaded in green highlights the possible final states after the optical
excitation. To assist the reader, a rectangle has been inserted to highlight the pseudogap of the AF phase DOS.}
\end{figure}

The phase-sensitivity of the transient reflectance of FeRh can be explained using the spin-polarized DOS for the AF and FM phases of FeRh. The total electronic density of states (DOS) for the AF and FM phases are displayed in Figure \ref{fig:AFDOS}. The occupied and unoccupied states following excitation are highlighted using a blue rectangle and yellow rectangle, respectively. As seen from the density of states (DOS) plot in Figure \ref{fig:AFDOS}a, the AF phase has a pseudogap extending from 0 to about 0.5 eV below the Fermi energy. The DOS of the occupied states that can be excited by 1.58 eV photons (yellow area in Figure \ref{fig:AFDOS}) matches well with the DOS in of the unoccupied states (green area). In the AF state, all bands are spin-degenerate such that the spin conservation of the optical transitions is not a limitation. As a result, the optical pump pulse is well absorbed. Thus, a decrease in $Delta$R(t)/R is observed in the AF phase. In the FM phase, the majority of the optically excitable states (-1.58 eV < E < 0 eV, yellow area in Figure \ref{fig:AFDOS}b) have spin-up polarization, while the majority of the unoccupied states are spin-down.
Since the optical transition induced by the pump pulse conserves spin, only few electrons can be excited. Thus, interband transitions are strongly suppressed in the FM phase and the pump pulse is only weakly absorbed. Consequently, a decrease in $\Delta$R(t)/R is not observed.\newline\indent The phase dependence of the electron-phonon coupling factor is supported by the electronic density of states of the AF and FM phases of FeRh. Electron scattering events occurring near the Fermi energy contribute most strongly to the electron-phonon coupling factor.\cite{unikandanunni_ultrafast_2022} The electronic DOS of the AF phase includes a pseudogap of approximately 0.5 eV about the Fermi energy in both spin channels. This pseudogap is not observed in the FM phase. This difference in the electronic DOS explains the changes in the electron-phonon coupling factor across the transition, and thus the differences in the electron-phonon thermalization times of the AF and FM phases.\section{\label{sec:level6}Conclusions\protect\\} We have used TDTR analysis to study the behavior of FeRh following optical pump pulse excitation at sample base temperatures below, above, and throughout the AF/FM transition. For these measurements, we used small temperature excursions to excite the system. Our results demonstrate the sensitivity of the transient reflectance to the magnetic order of the material. At low temperatures, when the majority of the sample is antiferromagnetic, the transient reflectance can be described by a significant negative transient signal followed by a slower negative signal. At temperatures above the transition region, the transient reflectance is positive and the fast transient signal observed in the AF phase was not detected. The change in the sign of the transient reflectance across the transition can be qualitatively explained by differences in the spin-polarized density of states within the energy range accessible to electrons affected by the pump pulse (-1.6 eV-1.6 eV). \newline\indent To explain the magnetic phase dependence of the fast transient signal, we used the two-temperature model in the weak perturbation regime. Using this phenomenological model, we determined that the phase sensitivity of the transient reflectance is caused by changes in the electronic heat capacity of the material and an increase in the electron-phonon coupling factor in the FM phase. Using the results of ab-initio calculations, the phase-dependent changes of these parameters can be explained by the differences in the electronic bandstructure near the Fermi energy, specifically the pseudogap observed in the electronic DOS of the AF phase that is absent in the electronic DOS of the FM phase. Conducting two-temperature model calculations using the NTMpy code and developing a linear systems model to fit the calculated electron and lattice temperatures, we determined the electron-phonon coupling factor in the AF phase. The results of this study demonstrate the tunability of the sign of the transient reflectance using the electronic bandstructure of the material. Although the AF phase is difficult to probe in the presence of FM domains, these results demonstrate a method for detecting the AF phase at temperatures when the FM phase is present. These results are significant, because they demonstrate the flexibility of TDTR to study the ultrafast response of magnetic materials which may otherwise be challenging to probe. 
\begin{acknowledgments}The TDTR data collection, characterization and analysis for this project was conducted at the University of Illinois at Urbana-Champaign (UIUC) using the Materials Research Lab facilities while  R. M. H. was a member of the research groups of Prof. David G. Cahill and Prof. Nadya Mason, whom we thank for guidance, fruitful discussions and support. This effort was continued by R.M.H. in the group of Prof. Thomas A. Searles supported by the National Science Foundation through the University of Illinois at Urbana-Champaign Materials Research Science and Engineering Center DMR-1720633 and in part by the U.S. National Science Foundation CAREER Award DMR-2047905. In addition, F. H. and A.C. would like to acknowledge support from the Director, Office of Science, Office of Science, Office of Basic Energy Sciences, Materials Sciences and Engineering Division, U.S. Department of Energy, under Contract No. DE-AC02-05-CH11231 within the Nonequilibrium Magnetic Materials Program (KC2204), which supported the growth and magnetic characterization of the materials studied in this paper. Further, additional support came from NSF DMR MRSEC and the Illinois Fellowship. V.U. and S.B. acknowledge support from the European Research Council, Starting Grant No. 715452 “MAGNETIC-SPEED-LIMIT.”
\end{acknowledgments}
%

\end{document}